\documentclass[%
 ,twocolumn%
,superscriptaddress%
,tightenlines%
,aps
,prd
]{revtex4}

\usepackage{bm}%
\usepackage{graphicx}

\begin{document}

\newcommand{\abs}[1]{\left| #1 \right|}
\newcommand{\nn}[0]{\nonumber \\ }
\def\rmin{r_{\text{min}}}
\def\rpz{R^{+/0}}
\def\aem{\alpha}

\title{Isospin violation in $e^+e^- \to B \bar B$}%

\author{Roland Kaiser}%
\email{kaiser@pauli.ucsd.edu}
\affiliation{University of California, San Diego\\
9500 Gilman Drive, La Jolla, CA 92093-0319}

\author{Aneesh V. Manohar}%
\email{amanohar@ucsd.edu}
\affiliation{University of California, San Diego\\
9500 Gilman Drive, La Jolla, CA 92093-0319}

\author{Thomas Mehen}%
\email{mehen@phy.duke.edu}
\affiliation{Department of Physics, Duke University, Durham, NC 27708}
\affiliation{Jefferson Laboratory, 12000 Jefferson Ave.,  Newport News, VA 23606\\}

\begin{abstract}
The ratio of the $B^+  B^-$ and $B^0\bar B^0$ production rates in $e^+ e^-$ annihilation is computed as a function of the $B$ meson velocity and $BB^*\pi$ coupling constant, using a non-relativistic effective field theory.
\end{abstract}

\date{August 2002}%
\maketitle

The dominant production mechanism for $B$ mesons at CLEO, BaBar and Belle is via the $P$-wave decay of the $\Upsilon(4S)$ state, $e^+ e^- \to \Upsilon(4S) \to B \bar B$. The final state can contain either charged ($B^+ B^-$) or neutral ($B^0 \bar B^0$) mesons, and the ratio of charged to neutral $B$ mesons produced enters many $B$ decay analyses, including studies of CP violation.
We will define  the ratio
\begin{eqnarray}
\rpz=1+\delta \rpz= {\Gamma \left( \Upsilon(4S) \to B^+ B^- \right) \over
\Gamma \left( \Upsilon(4S) \to B^0 \bar B^0 \right)},
\label{1}
\end{eqnarray}
which is unity in the absence of isospin violation. The experimental value measured by the BaBar Collaboration is $\rpz = 1.10 \pm 0.06 \pm 0.05$~\cite{babar}, and by the CLEO  collaboration is 
$1.04 \pm 0.07 \pm 0.04$~\cite{CLEO1} and $1.058 \pm 0.084 \pm 0.136$~\cite{CLEO2}.

Isospin violation is due to electromagnetic interactions, and due to the mass difference of the $u$ and $d$ quarks. In most cases, isospin violation is at the level of a few percent. However, it is possible that there can be significant isospin violation in $\Upsilon$ decay~\cite{atwood,lepage,byers}. The $\Upsilon(4S)$ is barely above $B \bar B$ threshold; the $B$ mesons are produced with a momentum $p_B \sim 338$~MeV and velocity $v/c=0.064$ [using $M_{\Upsilon(4S)}=10.58$~GeV, $M_{B} = 5.2792$~GeV], so that the final state is  non-relativistic. The electromagnetic contribution to $\rpz$ is a function of $v$ and the fine-structure constant $\aem$. In the non-relativistic limit, there are $1/v$ enhancements, and the leading contribution is a function of $\aem/v$,
\begin{eqnarray}
\rpz={ \pi \aem/v \over 1 - e^{-  \pi \aem/v}} \left(1 + {\aem^2\over 4 v^2} \right)=
1 + {\pi \aem \over 2v} + \mathcal{O}\left({ \aem^2 \over v^2} \right),
\label{2}
\end{eqnarray}
and can be obtained by solving the Schr\"odinger equation in a Coulomb potential for a $P$-wave final state~\cite{schiff}.
Corrections to this result are suppressed by powers of $\aem$ without any $1/v$ enhancements. For $\Upsilon(4S)$ decay, this gives $\rpz=1.19$, a significant enhancement of the charged/neutral ratio~\cite{atwood,lepage,byers}.

Lepage~\cite{lepage} computed corrections to Eq.~(\ref{2}) by assuming a form-factor at the meson-photon vertex, and found that $\delta \rpz$ could be significantly reduced from $0.19$, or even change sign. Recent advances in the study of heavy quark systems and non-relativistic bound states allow us to improve on this estimate of $\delta \rpz$. Since the final state $B$ mesons are non-relativistic, and have low momentum, the final state interactions of the $B$ meson can be treated using non-relativistic field theory combined with chiral perturbation theory~\cite{luke}.  At momentum transfers smaller than the scale of chiral symmetry breaking $\Lambda_\chi \sim 1$~GeV~\cite{amhg}, the photon vertex can be treated as pointlike. The $B$ and $B^*$ states have a mass splitting of $45.78\pm0.35$~MeV, which is small compared with the momentum $p_B$ of the $B$ meson, so the $B$ and $B^*$ must both be included in the effective theory. Since $p_B$ is much smaller than the mass of the $b$-quark, heavy quark spin symmetry holds~\cite{falk,Bc}, and one can treat the $B$ and $B^*$ as one multiplet described by the $H^{(b)}$ field of HQET~\cite{book}. Similarly, the $\bar B$ and $\bar B^*$ can be combined into a $H^{(\bar b)}$ field, whose properties are related to those of $H^{(b)}$ by charge conjugation~\cite{gjmsw}. At low velocities, the dominant isospin violation is that enhanced by factors of $1/v$, which is obtained by solving the Schr\"odinger equation with the
$H^{(b)}-H^{(\bar b)}$ interaction potential. The NRQCD counting rules~\cite{bbl} show that $B \bar B$ annnihilation is suppressed, and can be neglected.
 At low momentum transfer, the $H^{(b)}-H^{(\bar b)}$ potential is dominated by single-pion exchange. Isospin violation in the potential arises from Coulomb photon exchange, and from isospin violation in the pion sector due to the $\pi^+-\pi^0$ mass difference and $\eta-\pi^0$ mixing.

\begin{figure}
\includegraphics[width=5cm]{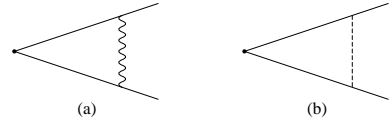}
\caption{One-loop correction to $\Upsilon(4S) \to B \bar B$ due to (a) photon and (b) pion exchange. \label{fig:1}}
\end{figure}

\begin{figure*}
\includegraphics[width=10cm]{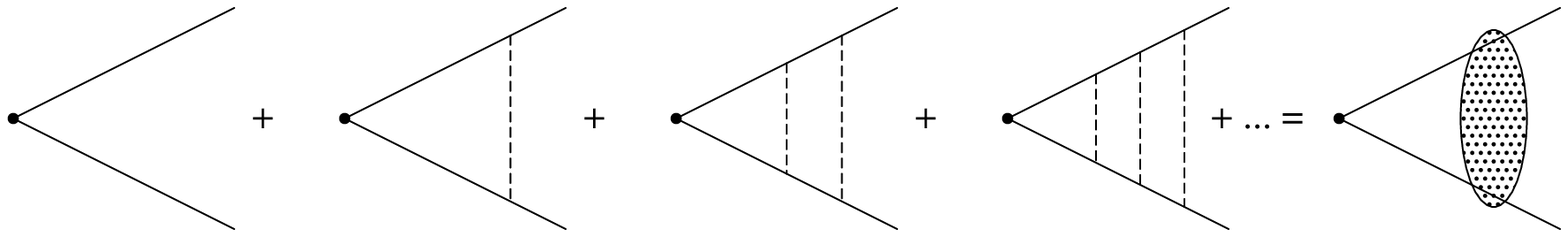}
\caption{Series of graphs summed by solving the Schr\"odinger equation. The dashed line represents pions and photons.
 \label{fig:2}}
\end{figure*}

In perturbation theory, the first contribution to $\delta \rpz$ is from the graphs in Fig.~\ref{fig:1}.
The one-loop photon graph gives the $\pi\aem/2v$ term in Eq.~(\ref{2}). It is enhanced by $\pi^2/v$ compared with a typical relativistic radiative correction, which is of order $\aem/\pi$, because of the non-relativistic nature of the integral. The one-loop pion graph is similarly enhanced by $\pi^2/v\sim 150$ compared with a typical chiral loop correction. As a result, the correction from Fig.~\ref{fig:1}(b) is not small, and cannot be treated in perturbation theory. However, it is possible to sum the  multiple pion exchanges by solving the Schr\"odinger equation using the one-pion  plus one-photon exchange potential. This sums the series of graphs shown in Fig.~\ref{fig:2}.
Additional corrections, such as vertex corrections,  are not included in the Schr\"odinger equation. However, these corrections are not enhanced by $\pi^2/v$, and so are subleading compared with the terms we have retained.

The $H^{(b)}-H^{(\bar b)}$ interaction potential is the same as the $H^{(b)}-H^{(b)}$ potential (by charge conjugation symmetry), and was computed in Ref.~\cite{exotic} which studied $bbqq$ exotic states. The potential depends on the $B^*B\pi$ coupling constant $g$ which is not known. Heavy quark symmetry implies that $g$ is the same as the $D^*D\pi$ coupling. The $D^*$ can decay into $D\pi$ (via the coupling $g$) or $D \gamma$ (via electromagnetic interactions), and the decay rates can be used to obtain $g$~\cite{amundson,cho}. A fit to the experimental data gives two possible solutions, 
$g=0.27\,^{+0.04}_{-0.02}\, {}^{+0.05}_{-0.02}$ or $g=0.76\,^{+0.03}_{-0.03}\,  {}^{+0.2}_{-0.1}$~\cite{is}, with the smaller value being preferred. 
A recent measurement of the $D^{*+}$ width by the CLEO collaboration gives $g=0.59 \pm 0.01 \pm 0.07$~\cite{gCLEO}. We will give our results as a function of $g$.

The $\Upsilon(4S)$ is a $1^{--}$ state, and can decay into five possible channels, (i) $B \bar B$ with $S=0$, $\ell=1$, (ii) $B^*\bar B^*$ with $S=0$, $\ell=1$, (iii) $B^*\bar B^*$ with $S=2$, $\ell=1$, (iv) $B^*\bar B^*$ with $S=2$, $\ell=3$ and (v) $B\bar B^*+B^*\bar B$ with $S=1$, $\ell=1$, where $\ell$ is the orbital angular momentum and $S$ is the total spin. Since the $\Upsilon(4S)$ is below $BB^*$ and $B^*B^*$ threshold, only the first state is allowed as a final state, but all five states need to be included as intermediate states in the calculation. [The actual number of states is double this, since one has both charged and neutral channels.] Let $\eta,\beta=1-5$ denote one of the five possible $\ell S$ states and $a,b=1,2$ denote
the charged and neutral sectors, respectively, so that a given channel is labeled by the index pairs $\eta a$ or $\beta b$.
The radial Schr\"odinger equation  has the potential
\begin{eqnarray}
V^\pi_{\eta a, \beta b}(r) +V^\gamma_{\eta a, \beta b}(r) +V^\ell_{\eta a, \beta b}(r)
+ M_{\eta} \delta_{\eta\beta}\delta_{a,b}
\label{3}
\end{eqnarray}
where $V^\pi$ is the pion potential, $V^\gamma$ is the Coulomb potential, $V^\ell$ is the angular momentum potential, and
$M_\eta$ is the contribution due to the $B^*-B$ mass difference, $\Delta m$, 
\begin{eqnarray}
&&M_1=0,\qquad M_5=\Delta m, \nn
&& M_2=M_3=M_4=2\Delta m.
\label{4}
\end{eqnarray}
The $B^0-B^+$ mass difference is $0.33\pm 0.28$~MeV~\cite{pdg}, and will be neglected in our analysis. Note that a  $B^0-B^+$ mass difference of $0.33$~MeV contributes about $0.05$  to $\delta\rpz$ from the $p^3$ dependence of the phase space of the $P$-wave decay.

The angular momentum potential is
\begin{eqnarray}
V^\ell_{\eta a, \beta b}(r) &=& {\ell_\eta (\ell_\eta+1) \over m_B r^2} 
\delta_{a,b} \delta_{\eta \beta}
\label{5}
\end{eqnarray}
where $\ell_\eta=(1,1,1,3,1)$ are the angular momenta of the various channels. The denominator is $m_B$ since $m_B/2$ is the reduced mass of the $B\bar B$. The Coulomb potential is
\begin{eqnarray}
V^\gamma_{\eta a, \beta b}(r) &=& -{\aem \over r} \delta_{\eta\beta} \delta_{a1}\delta_{b1},
\label{6}
\end{eqnarray}
where $\aem$ is the fine-structure constant. It only contributes to the charged sector $a=b=1$, and does not mix different $\ell S$ states. 

The pion potential can be computed using the techniques given in Ref.~\cite{exotic,largen,ej}:
\renewcommand{\arraystretch}{1.5}
\arraycolsep 3pt
\begin{eqnarray}
V^\pi_{\eta a, \beta b}(r) &=& \left[
\begin{array}{cc}
h_+^2 U_{\eta\beta}\left(m_{\pi^0},r\right) & 2 U_{\eta\beta}\left(m_{\pi^+},r\right) \\
2 U_{\eta\beta}\left(m_{\pi^+},r\right)& h_0^2 U_{\eta\beta}\left(m_{\pi^0},r\right)
\end{array}
\right]_{ab}
\label{7}
\end{eqnarray}
where $h_+^2=1.01$, $h_0^2=0.99$~\cite{ej}.
The structure of the potential is easy to understand. The off-diagonal elements are transition amplitudes between the charged and neutral sectors due to $\pi^+$ exchange, and depend on the charged pion mass $m_{\pi^+}$ and coupling constant $g$, which is included in the definition of $U$. The diagonal matrix elements are due to $\pi^0$ exchange, and depend on $m_{\pi^0}$. In the absence of $\eta-\pi^0$ mixing, the $\pi^+$ coupling constant is $\sqrt{2}$ time the $\pi^0$ coupling which gives the $2:1$ ratio of the off-diagonal to diagonal elements. The values of $h_+$ and $h_0$ differ from unity due to $\eta-\pi^0$ mixing~\cite{ej}. 

The computation of the matrix $U_{\eta\beta}(m_\pi,r)$ is non-trivial. The answer is that
\begin{eqnarray}
U(m,r) = T\,  \widetilde U(m ,r)\, T^t
\label{8}
\end{eqnarray}
where
\renewcommand{\arraystretch}{1.0}
\begin{eqnarray}
\tilde U(m,r) &=& {g^2 m^2 e^{-m r} \over 8 \pi f^2 r}
\left[ \begin{array}{ccccc}
1 & 0 & 0 & 0 & 0 \\
0 & u_1 & 0 & 0 & 0 \\
0 & 0 & u_2 & 0 & 0 \\
0 & 0 & 0 &u_2 & 0 \\
0 & 0 & 0 &  0 &u_1 
\end{array}\right] ,\nn
u_1(m,r) &=& \left(1 + {2 \over m r} \right)^2,\nn
u_2(m,r) &=& -\left(1 + {2 \over m r} + {2 \over m^2 r^2} \right),
\label{9}
\end{eqnarray}
$f\sim 132$~MeV is the pion decay constant
and
\renewcommand{\arraystretch}{1.75}
\begin{eqnarray}
T=\left( \begin{array}{ccccc}
{1\over 2} & -{1\over 2}&  0& {1 \over \sqrt 2} & 0 \\
{\sqrt 3 \over 2} & {1 \over 2 \sqrt 3} & 0 & - {1 \over \sqrt 6 } & 0\\
0 & -{2 \over \sqrt {15}} & -\sqrt{3 \over 10} & - \sqrt{2 \over 15} & -\sqrt{3 \over 10}\\
0 & \sqrt{2 \over 5} & - {1 \over \sqrt 5}  & {1 \over \sqrt 5}  & -{1 \over \sqrt 5}  \\
0 & 0 & - {1 \over \sqrt 2} & 0 & {1 \over \sqrt 2}
\end{array}\right),
\label{10}
\end{eqnarray}
and $T^t$ is the transpose of $T$.

Equation~(\ref{7})  is the leading contribution to the long distance part of
the potential. As argued in Ref.~\cite{exotic}, Eq.~(\ref{7})
will dominate the potential until $r \sim 1/(2 m_\pi)$ at which point two-pion exchange begins to contribute. We introduce a cutoff $\rmin=1/(2m_\pi)$, and use Eq.~(\ref{7}) for $r \ge \rmin$, and set $V^\pi=0$ for $r < \rmin$. The short distance part of the potential can be included into a renormalization of the production vertex. The Coulomb potential will be allowed to act until $r=0$.

The $\Upsilon(4S)$ is produced by the space component of the electromagnetic current $\bar b \gamma^i b$. Heavy quark spin symmetry holds in the $\Upsilon$ system~\cite{falk,Bc}, so the
$\Upsilon(4S)$ decays into $H^{(b)} -H^{(\bar b)}$ such that the spins of the heavy quarks in the final mesons are combined to form the spin of the $\Upsilon(4S)$, i.e.\ the polarization of the virtual photon. The orbital angular momentum and spin of the light degrees of freedom are combined to form total angular momentum zero. A little Clebsch-Gordan algebra shows that the amplitude for the $\Upsilon(4S)$ to decay into the five channels is~\cite{falk}
\begin{eqnarray}
A _{\eta a}= c_a \left( {1\over 2 \sqrt 3}, -{1\over 6}, {\sqrt{5}\over 3}, 0,
 -{1\over \sqrt 3} \right)_\eta \ .
\label{11}
\end{eqnarray}
The amplitude for decay to the $\ell=3$ channel is zero to this order in the velocity expansion. The coefficients $c_a,\ a=1,2$ are unknown, but the absolute values of $c_a$ are irrelevant for the computation of $\rpz$; all that is needed is the ratio $c_1/c_2$ of the charged to neutral production amplitudes. The dominant production of the $B$ mesons is via the isosinglet $\Upsilon(4S)$ state, in which case $c_1=c_2$.  Isospin violating effects, including direct production of $B$'s not via the $\Upsilon(4S)$ lead to a deviation of $c_1/c_2$ from unity. As discussed above, cutoff effects in the potential can be absorbed into the production amplitudes $c_a$.  One expects short-distance corrections to introduce isospin violation in the ratio $c_1/c_2$ of a few percent, the typical size of other isospin violating effects in hadron physics.  We will  define $\delta c$ by $c_1/c_2 = 1 + \delta c$. The value of $\delta c$ is related to the value of $\rmin$, since changes in the cutoff induce changes in the Lagrangian coefficients. Since $\delta c$ is unknown, our computation of $\rpz$ is uncertain at the 5\% level; however the uncertainity is much smaller than the expectation that $\delta \rpz$ is 19\% from Coulomb interactions alone. Cutting off the Coulomb potential at short distances reduces the value of $\delta \rpz$. Since the Coulomb potential is the dominant source of isospin violation, one expects that $\delta c$ will be negative.

The method of computation is as follows. One solves the Schr\"odinger equation with potential Eq.~(\ref{7}). The boundary condition on the wavefunction as $r \to \infty$ is that one has a plane wave plus an outgoing scattered wave. 
[One can see this directly from the sum of graphs in Fig.~\ref{fig:2}.] Only the $B^+ B^-$ and $B^0 \bar B^0$ states exist as propagating modes as $r \to \infty$; the other channels have exponentially decaying wavefunctions. The plane wave state is chosen to be in the $B^+ B^-$ or  $B^0\bar B^0$ channels to compute the charged or neutral meson production rates, respectively. The overlap of the computed wavefunction as $r\to 0$ with the production amplitude
 Eq.~(\ref{11}) gives the final production amplitude, the absolute square of which
gives the production rate. [Note that the wavefunction near $r=0$ can have all five channels.]
The answer for $\rpz$ depends on $\delta c$, $g$ and the velocity $v$ of the outgoing $B$ meson. 
Provided the dominant production mechanism is via the photon coupling to the heavy quark, the result for $\rpz$ holds even away from the $\Upsilon(4S)$ resonance since it depends only on the quarks being non-relativistic. The value of $c_a$ will depend strongly on the beam energy, and peak at the resonance, but $\delta c$, the isospin violation in the production amplitude should be a smooth function of energy.

\begin{figure}[tbp]
\includegraphics[width=6cm]{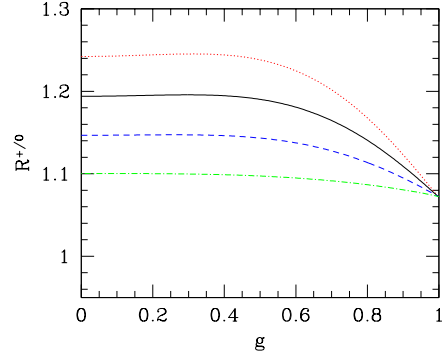}
\vspace{-35pt}
\caption{$\rpz$ as a function of $g$ for $v=0.064$ and  $\delta c=0.02$ (dotted), $0.0$ (solid), $-0.02$ (dashed) and $-0.04$ (dot-dashed).
\label{plot:1}}
\end{figure}

In Fig.~\ref{plot:1} we have plotted $\rpz$ as a function of $g$ for  $\delta c $ and $v=0.064$, the value in $\Upsilon(4S)$ decay.
 $\rpz$ is approximately constant and equal to its value from only Coulomb corrections, Eq.~(\ref{7}), until $g>0.6$, at which point $\rpz$ starts to decrease. $\rpz$ is approximately constant for small $g$ even though the shifts in the production amplitudes are large. The one-loop pion graph in Fig.~\ref{fig:1} is about three times the tree-level graph. Summing the pion graphs in Fig.~\ref{fig:2} gives about a 20\% (for $g \sim 0.6$) shift in the charged and neutral production rates. The rates into the charged and neutral channels vary by about a factor of two for the range of Yukawa couplings in Fig.~\ref{plot:1}, but their ratio $\rpz$ varies by  about 10\%. For larger values of $g$ than those shown, $\rpz$ has rapid $v$ dependence due to the formation of meson bound states, because the pion-exchange potential is sufficiently attractive. For our choice of parameters, this occurs for $g\sim 1.3$, well outside the allowed range~\cite{is,gCLEO}.

\begin{figure}
\includegraphics[width=6cm]{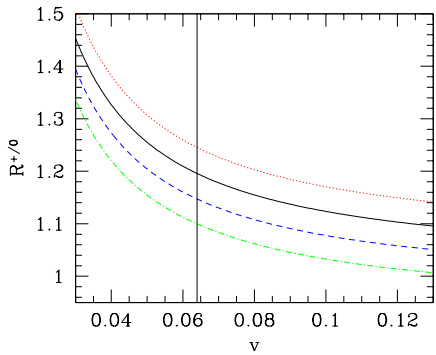}
\vspace{-35pt}
\caption{$\rpz$ as a function of $v$ for $g=0.3$ and  $\delta c=0.02$ (dotted), $0.0$ (solid), $-0.02$ (dashed) and $-0.04$ (dot-dashed).
\label{plot:2}}
\end{figure}

\begin{figure}
\includegraphics[width=6cm]{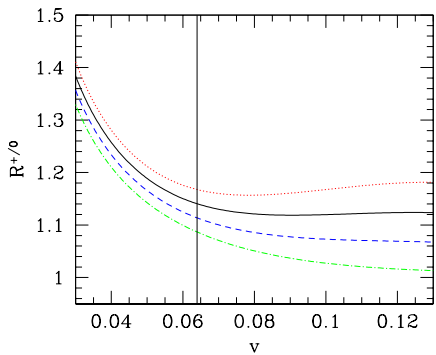}
\vspace{-35pt}
\caption{$\rpz$ as a function of $v$ for $g=0.8$ and  $\delta c=0.02$ (dotted), $0.0$ (solid), $-0.02$ (dashed) and $-0.04$ (dot-dashed).
 \label{plot:3}}
\end{figure}

In Fig.~\ref{plot:2} and \ref{plot:3}, we have plotted $\rpz$ as a function of  velocity for different values of $\delta c$ for two illustrative choices $g=0.3$ and $g=0.8$ consistent with the two solutions for $g$ found in Ref.~\cite{is}. The vertical line is the velocity at the $\Upsilon(4S)$.
At the $\Upsilon(4S)$ peak,
for $g=0.8$, $\rpz$ varies from $1.17$ to about $1.09$, whereas for $g=0.3$, $\rpz$ varies between about $1.25$ and $1.1$.

\begin{figure}
\includegraphics[width=6cm]{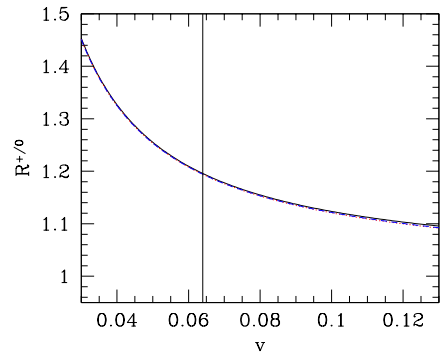}
\vspace{-35pt}
\caption{$\rpz$ as a function of $v$ for $g=0.3$, $\delta c=0$, and cutoffs $\rmin=1/(2m_\pi)$ 
(solid), $1/\sqrt 2 m_\pi$ (dashed), $1/m_\pi$ (dotted).
 \label{plot:4}}
\end{figure}

\begin{figure}
\includegraphics[width=6cm]{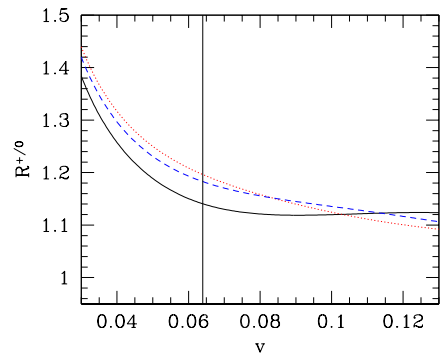}
\vspace{-35pt}
\caption{$\rpz$ as a function of $v$ for $g=0.8$, $\delta c=0$,  and cutoffs $\rmin=1/(2m_\pi)$ 
(solid), $1/\sqrt 2 m_\pi$ (dashed), $1/m_\pi$ (dotted).
 \label{plot:5}}
 \end{figure}

In Figs.~\ref{plot:4} and \ref{plot:5}, we have plotted $\rpz$ as a function of $v$ for $g=0.3$ and $g=0.8$, respectively, for different values of the cutoff from $\rmin=1/(2m_\pi)$ to $1/m_\pi$.  For small values of $g$, the variation of the cutoff does not change $\rpz$. For larger values of $g$, the cutoff variation is consistent with expectations from naive dimensional analysis~\cite{amhg}. A factor of two variation in the cutoff introduces a $ 4\%$ variation in $\rpz$.

The absolute value of $\rpz$ depends on the value of $\delta c$, and the cutoff $\rmin$. If $g$ is small ($\sim 0.3$, the preferred value in Ref.~\cite{is}),  then for values of $\delta c$ consistent with expectations from dimensional analysis, one expects $\delta \rpz \agt  0.1$. The Yukawa corrections do not significantly change $\rpz$ from the Coulomb value. We note, however, that this is due to a cancellation in $\rpz$ after summing the graphs in Fig.~\ref{fig:2}; the one loop pion correction from Fig.~\ref{fig:1} is about three, and is not small.
If $g$ is close to the larger value $g=0.8$, then $\delta\rpz$ at the $\Upsilon(4S)$ is smaller, but still around $0.1$. In this case, there is some cutoff dependence, so $\rpz$ is more uncertain.

The dependence of $\rpz$ on $v$ (or equivalently, $\sqrt{s}$) is calculable. One can see that the curves in Fig.~\ref{plot:2} have a different shape from those in Fig.~\ref{plot:3},
so measuring $\rpz$ as a function of $v$ can provide information on the $B^*B\pi$ and $D^*D\pi$ coupling $g$, which is needed for many calculations, such as the ratio of the $B_s-\bar B_s$ to $B -\bar B$ mixing amplitudes~\cite{gjmsw}.

\acknowledgments

We would like to thank S.~Fleming and V.~Sharma for discussions.
This work was supported in part by the Department of Energy under grants DOE-FG03-97ER40546,
DE-FG02-96ER40945 and DE-AC05-84ER40150. RK was supported by Schweizerischer Nationalfonds.

\end{document}